\documentclass[prl,twocolumn,preprintnumbers,nofootinbib,superscriptaddress,showpacs]{revtex4}
  
\usepackage{amsmath,amssymb}
\pdfoutput=1
\usepackage{epsfig}  
\usepackage{graphicx}               
\usepackage{url}
\usepackage{hyperref}

\usepackage{color}
  
\newcommand{\nc}{\newcommand}  

\nc{\beq}{\begin{equation}}  
\nc{\eeq}{\end{equation}}  
\nc{\beqa}{\begin{eqnarray}}  
\nc{\eeqa}{\end{eqnarray}}  
\nc{\bea}{\begin{eqnarray}}  
\nc{\eea}{\end{eqnarray}}  
\nc{\ra}{\rightarrow}  
\nc{\lsim}{\begin{array}{c}\,\sim\vspace{-21pt}\\< \end{array}}  
\nc{\gsim}{\begin{array}{c}\sim\vspace{-21pt}\\> \end{array}}  
\nc{\Tr}{{\rm Tr}}
\nc{\slsh}{\slash\hspace*{-0.22cm}}

\def\be{\begin{equation}}
\def\ee{\end{equation}}
\def\bea{\begin{eqnarray}}
\def\eea{\end{eqnarray}}
\def\bit{\begin{itemize}}
\def\eit{\end{itemize}}

\begin{document}
\title{Geometric Compatibility of IceCube TeV-PeV Neutrino Excess and its Galactic Dark Matter Origin}
\author{Yang Bai}
\affiliation{Department of Physics, University of Wisconsin, Madison, WI 53706, USA}
\author{Ran Lu}
\affiliation{Department of Physics, University of Wisconsin, Madison, WI 53706, USA} 
\affiliation{Michigan Center for Theoretical Physics, University of Michigan, Ann Arbor, MI 48109, USA}
\author{Jordi Salvado}
\affiliation{Department of Physics, University of Wisconsin, Madison, WI 53706, USA} 
\affiliation{Wisconsin IceCube Particle Astrophysics Center, Madison, WI 53706, USA}


\begin{abstract} 
We perform a geometric analysis for the sky map of the IceCube TeV-PeV neutrino excess and test its compatibility with the sky map of decaying dark matter signals in our galaxy. Using both Kolmogorov-Smirnov and the likelihood-ratio tests, we have found that the observed event sky map prefers to have a combination of the galactic dark matter and a homogeneous background contributions, compared to a purely galactic dark matter origin. For the assumption that the galactic dark matter is responsible for all neutrino excess, the current data can also exclude a wide range of dark matter profiles except flatter profiles such as the isothermal one. We also consider several representative decaying dark matter spectra, which can provide a good fit to the observed spectrum at IceCube with a dark matter lifetime of around 12 orders of magnitude longer than the age of the universe.   
\end{abstract}

\maketitle

\paragraph{\bf Introduction}
\label{sec:intro}
One of the important tasks for physicists is to understand the nature of dark matter (DM). The indirect search of DM from its self-annihilation or decay serves as a promising approach to learn additional interactions of DM with Standard Model (SM) particles. Among many potential products from DM annihilation or decay, neutrino serves a useful candidate because its propagation is less disturbed by the interstellar medium and its observation may point out the DM geometric distribution in the galaxy. 

The existing searches of DM from cosmic neutrinos have been concentrated on the galactic center, dwarf galaxies, clusters of galaxies or the center of the Sun~\cite{Halzen:2009vu,Tanaka:2011uf,Aartsen:2012kia,Adrian-Martinez:2013ayv}. All previous searches have found good agreements between the neutrino spectrum and the predicted astrophysical background. The story has been changed recently from the observation of 28 high energy neutrino events at IceCube for neutrino energy above around  30 TeV~\cite{IceCube:science}, which is well above the predicted number of the background events, $10.6^{+5.0}_{-3.6}$~\cite{Honda:2006qj,Enberg:2008te} and has a $4.0\sigma$ inconsistence with the standard atmospheric backgrounds. 

This IceCube result is based on data taken between May 2010 and May 2012 using detectors with 79 and 86 strings respectively and has a total integrated time of 662 days. The observed 28 events have two events slightly above 1 PeV~\cite{Aartsen:2013bka} and the remaining 26 events with an energy between 25 TeV and 0.3 PeV. The observed events can also be divided into ``track" and ``cascade" events, depending on event shapes. The track events are most likely produced by muon neutrinos via charge-current interactions, while the cascade events could come from electron neutrinos with charge-current interactions or all types of neutrinos with neutral-current interactions. Among the 28 events, the seven track events have a good angular resolution with around $1^\circ$ uncertainty around the event direction, while the other 21 cascade events have poor angular resolutions ranging from $\sim10^\circ$ to $\sim50^\circ$. 

The angular resolutions of those events play an important role for identifying the geometric origin of the neutrino excess. The IceCube collaboration has performed a point source analysis for the 28 events and found that there is no significant evidence of spatial clustering and the $p$-value for the hypothesis of a uniform event distribution is 80\%~\cite{IceCube:science}. Curious about the possible linkage between the TeV-PeV neutrino excess at IceCube and the mysterious DM in our universe, in this paper we analyze the IceCube data with a special attention on its geometric distributions, and study the statistical significance of its potential DM origin, which prefers to have more signal events around the galactic center because of the DM spatial profile. 

One could consider DM annihilation as an explanation. However, due to the unitarity bound~\cite{Griest:1989wd,Hui:2001wy}, we found that the annihilation rate for a DM mass around one PeV is about four orders of magnitude lower than the required one for the IceCube data. Therefore, we concentrate on the decaying DM case, which can match to the required rate for a DM lifetime of $10^{28}-10^{29}$~s. In this paper we do not provide a theoretical understanding of the DM mass scale and the decay lifetime, but we want to point out that a heavy DM with a non-thermal history has been widely predicted in many models~\cite{Chung:1998zb,Chung:1998ua,Covi:2009xn,Feldstein:2013kka}. 


Before entering into our detailed geometric analysis, we point out other recent explanations for the IceCube neutrino excess including cosmogenic productions via photo-meson interactions~\cite{Cholis:2012kq,Laha:2013lka,Anchordoqui:2013qsi,Winter:2013cla}, galactic sources~\cite{Gonzalez-Garcia:2013iha}, active galactic nuclei~\cite{Kalashev:2013vba,Stecker:2013fxa,Winter:2013cla}, gamma-ray bursts~\cite{Murase:2013ffa,Winter:2013cla}, and a leptoquark beyond the SM~\cite{Barger:2013pla}.

\paragraph{\bf Geometric analysis for decaying dark matter based on the Kolmogorov-Smirnov test}
\label{sec:geometry}

%
\begin{figure*}[th!]
\begin{center}
\includegraphics[width=0.47\textwidth]{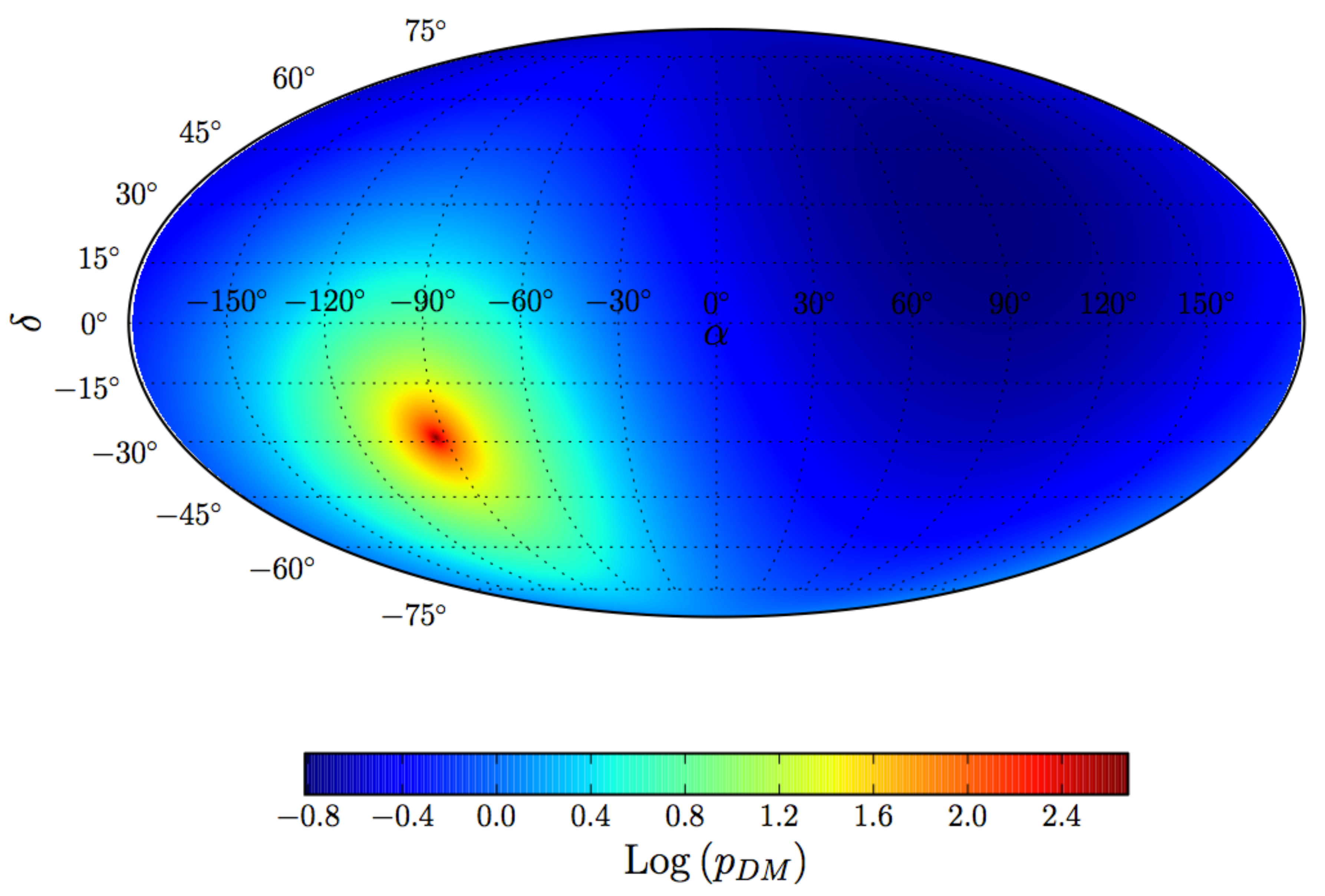} 
\hspace{0.3cm}
\includegraphics[width=0.47\textwidth]{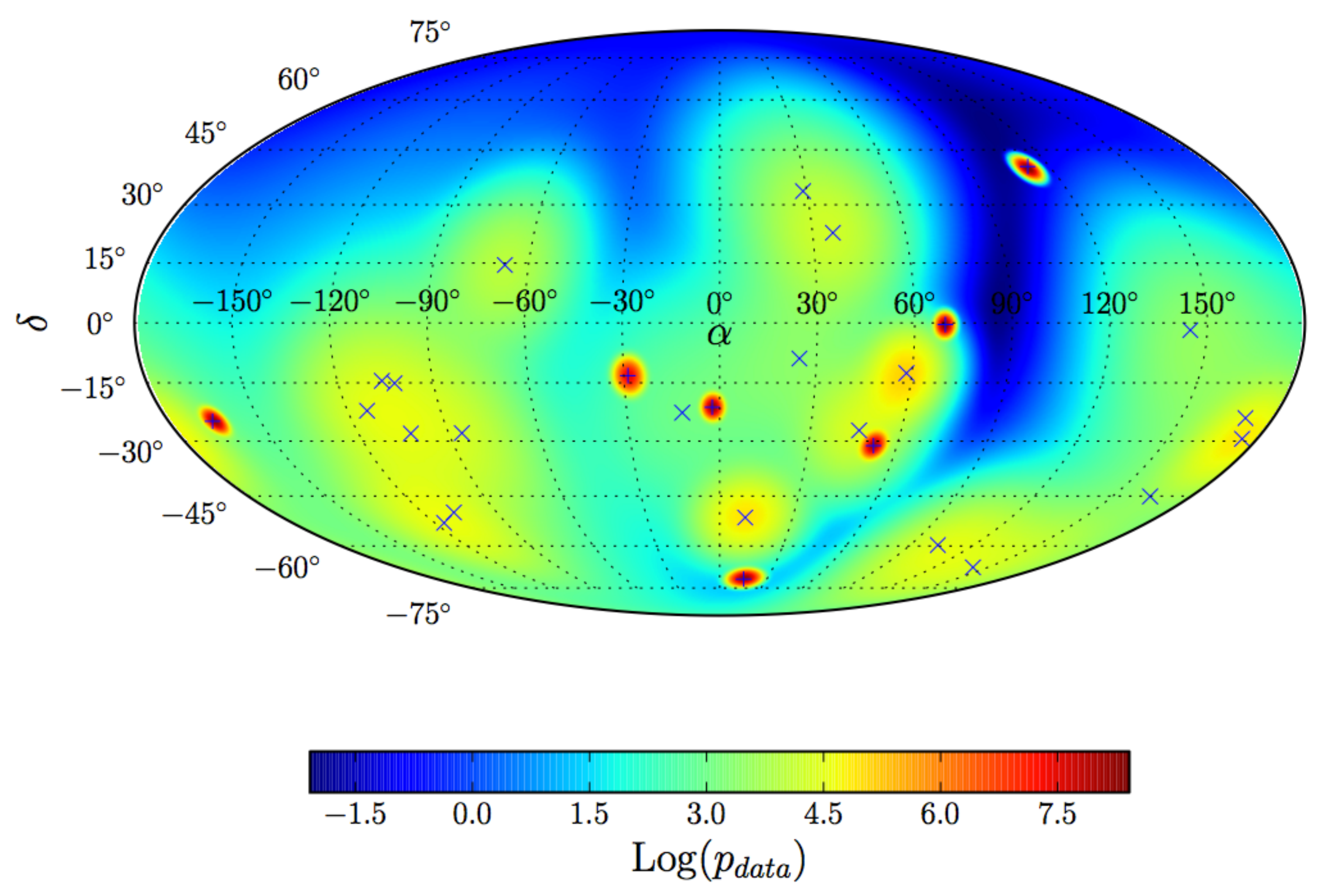}
\caption{Left panel: the sky map of the neutrinos from decaying DM with an Einasto profile in Eq.~(\ref{eq:einasto}). Right panel: the sky map of the IceCube 28 events after taking into account the angular resolution. The seven red spots correspond to the seven ``track" events.}
\label{fig:sky-map}
\end{center}
\end{figure*}

Our main goal is to study the compatibility of the neutrino sky map from DM and the sky map of the observed events at IceCube. The signal distribution from DM decays depends on the DM spacial profile in our galaxy. For the Einasto profile~\cite{Graham:2006ae}, one has
\beqa
\rho_{\rm DM}(r) = \rho_{\odot} \, e^{-\frac{2}{\bar\alpha}\left[ \left( \frac{r}{r_s} \right)^{\bar\alpha} - \left( \frac{r_{\odot}}{r_s} \right)^{\bar\alpha} \right]  } \,,
\label{eq:einasto}
\eeqa
with $r_s=20$~kpc and $\bar{\alpha}=0.17$ as a standard choice. A larger value of $\bar{\alpha}$ has a flatter DM profile.
Here, $r_\odot \simeq 8.5$~kpc is the distance from the Sun to the galactic center; $\rho_\odot \simeq 0.3$~GeV~cm$^{-3}$ is the approximate DM density in the solar system. The neutrino signal from DM decay is calculated by the line-of-sight integral along a given direction~\cite{Covi:2009xn}
\beqa
\frac{d\Phi_\nu}{dE_\nu\,db\,dl} = \frac{dN}{N dE_\nu} \, \frac{1}{\tau_{\rm DM}\, m_{\rm DM}} \frac{\cos{b}}{4\pi} \int ds \,\rho_{\rm DM}[r(s)] \,,
\eeqa
where the integral of $s$ is along the line of sight and the relation between $r$ and $s$ is $r^2 = s^2 + r^2_\odot - 2 s \,r_\odot \cos{l}\cos{b}$, where $- 90^\circ \leq b < 90^\circ$ and $- 180^\circ \leq l < 180^\circ$ as the latitude and longitude angles in the galactic coordinate. $\tau_{\rm DM}$ is the DM lifetime and $m_{\rm DM}$ is the DM mass. The normalized neutrino differential spectrum is $dN/(N d E_\nu)$. The integrated neutrino flux from DM is
\beqa
\Phi_\nu =1.7\times 10^{-12}~\mbox{cm}^{-2}\,\mbox{s}^{-1}\,\mbox{sr}^{-1}\times  \frac{10^{28}~s }{\tau_{\rm DM} } \times \frac{1~\mbox{PeV}}{m_{\rm DM} }  \,.
\eeqa
For the integrated time of $662$~days and 10 $\mbox{m}^2\cdot \mbox{sr}$ acceptance area for the energy around 100 TeV, there could be around 10 events observed at IceCube.

The geometric distribution of the IceCube events is represented in the equatorial coordinate. We, therefore, translate the DM generated event distribution from the galactic coordinate in  the latitude and longitude angles $(b, l)$ to the equatorial coordinate in the declination angle and the right ascension angle $(\delta, \alpha)$ (see Ref.~\cite{Neunhoffer:2004ha} for details). We define the DM probability distribution using the normalized flux
\beqa
p_{\rm DM}(\delta,\alpha)=\frac{1}{\Phi_\nu}\frac{d\Phi_\nu(\delta,\alpha)}{d\delta\,d\alpha} \,,
\eeqa
with the DM event sky map shown in the left panel of Fig.~\ref{fig:sky-map}. For all or subsets of the observed 28 events from IceCube, we construct the data probability distribution using the solid-angular error $\sigma_i$ for each event by assuming a Gaussian distribution
\beqa
p^{N\,{\rm events}}_{\rm data}(\delta,\alpha)=\frac{1}{N}\sum_{i \in N} \frac{1}{2\pi\sigma_i^2} \exp\left[-\frac{{\Delta R}(\delta_i, \alpha_i; \delta,\alpha)^2}{2\pi\sigma_i^2}\right],
\eeqa
where $\Delta R(\delta_i, \alpha_i; \delta,\alpha)$ is the angular distance between the points $(\delta_i, \alpha_i)$ and $(\delta, \alpha)$ on the sphere. In the right panel of Fig.~\ref{fig:sky-map}, we show the sky map of the observed $N=28$ events at IceCube after implementing the angular resolution for each event. Comparing these two maps, one can see that both have a concentration of events around the galactic center direction. On the other hand, the DM sky map has very few events in the right and upper corner, while the IceCube data map has some population in this region. 

\begin{figure*}[th!]
  \begin{center}
    \hspace*{-0.0cm}
    \includegraphics[width=0.45\textwidth]{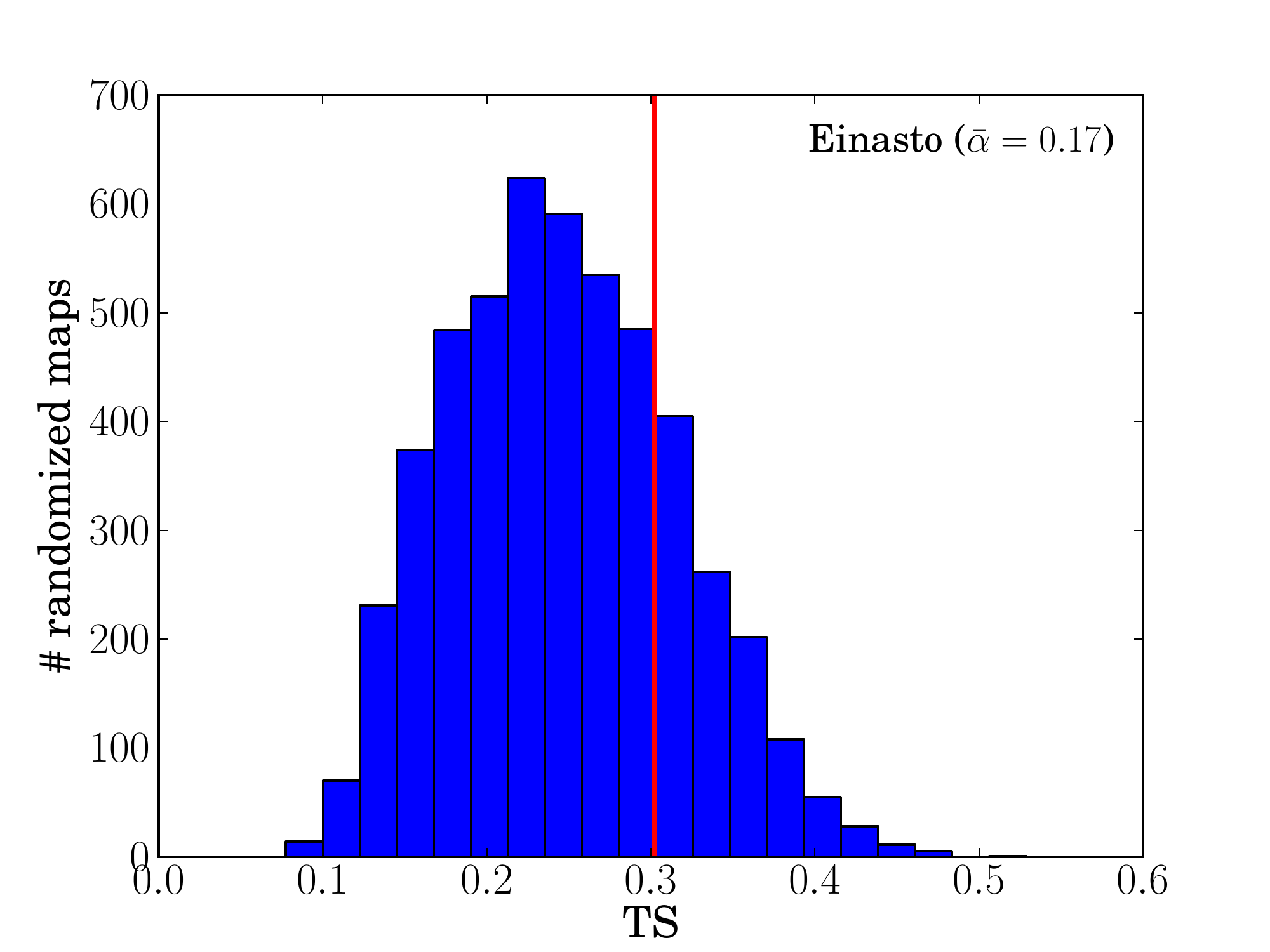} 
    \hspace{2mm}
    \includegraphics[width=0.45\textwidth]{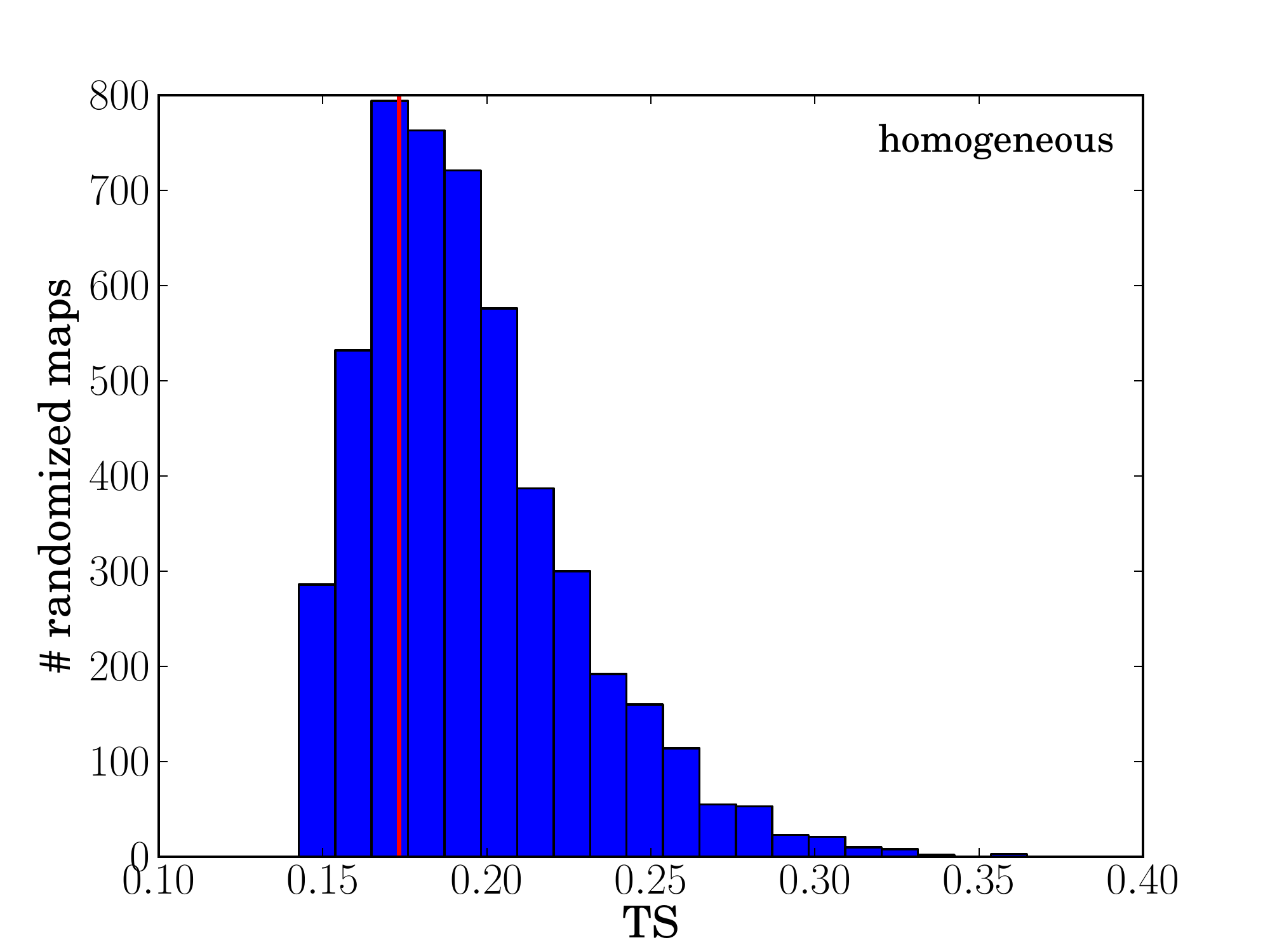}
    \caption{Left panel: the TS distribution for the Einasto model with $\bar \alpha=0.17$ ($p_{\rm value}=21.98\%$) against random sky maps with random right-ascension angles for the 28 events. 
Right panel: the TS distribution for a homogeneous distribution ($p_{\rm value}=72.14\%$). The TS value for the data is shown on the red lines.
}    \label{fig:TS-histograms}
  \end{center}
\end{figure*}

To quantify the similarity of the two sky maps in Fig.~\ref{fig:sky-map}, we perform a statistical test to calculate the $p$-value of the hypothesis of decaying DM as an explanation of IceCube neutrino excess. We first use a two-dimensional version of the Kolmogorov-Smirnov (KS) test statistics (TS)~\cite{Peacock1983MNRAS.202..615P} to study the compatibility between the data and the DM hypothesis. We will use the maximum likelihood-ratio test as well later. The KS test statistics is defined as the largest absolute difference between cumulative probability distributions of the data and the model. It takes better account of  the relation among data points than the traditional likelihood-ratio test.

To make the definition of the TS less sensitive to the integration directions, we consider a set of four possible integration regions,
\beqa
S(\delta_0, \alpha_0)&=&\left\{(\delta<\delta_0,\alpha<\alpha_0),\;(\delta>\delta_0,\alpha<\alpha_0), \right.
\nonumber \\
&&  \left. \;
(\delta<\delta_0,\alpha>\alpha_0), \; (\delta>\delta_0,\alpha>\alpha_0)\right\}\,,
\eeqa 
for a given boundary choice $(\delta_0, \alpha_0)$. The TS or the difference of the cumulative probability distributions is defined by
\beqa
&&\hspace{1cm}{\rm TS}(\delta_0, \alpha_0)\equiv \\
&&\sup_{r\in S(\delta_0, \alpha_0)} \left|\int_{r}d\delta\, d\alpha\, p_{\rm model}(\delta, \alpha) 
-\int_{r} d\delta\, d\alpha\, p_{\rm data}(\delta, \alpha)\right|\, .  \nonumber 
\eeqa 
Choosing the largest value for all possible boundary choices, we have the KS test statistics as
\beqa
{\rm TS}_{\rm KS}=\sup \left\{ \bigcup_{\rm{all}\,(\delta_0, \alpha_0)} \mbox{TS}(\delta_0, \alpha_0)\right\}\,.
\eeqa  

To calculate the $p$-value for the decaying DM as an explanation for the data, we generate random event maps by choosing a random (according to the model profile) right-ascension angle but keeping the same inclination angle and resolution of the event in the data. 
In the left panel of Fig.~\ref{fig:TS-histograms}, we show the TS distribution of the reference decaying DM model against maps of randomly sampled 28 events. The red vertical line indicates the test statistics TS(DM) of DM against the observed 28 events at IceCube. The $p$-value, or the probability of having TS(DM) smaller than the TS value from a random event map, is 21.98\% for the Einasto model with $\bar\alpha=0.17$. To test how good the observed 28 events agree with a homogeneously geometrical distribution, we perform the same calculation by assuming a homogeneous model (in the right panel of Fig.~\ref{fig:TS-histograms}) 
and found that the $p$-value for a homogeneous distribution is 72.14\% for all 28 events.

Since the atmospheric backgrounds are dominated in lower energies~\cite{Honda:2006qj,Enberg:2008te}, a bigger fraction of the observed events could be from DM signals if only relatively high energy events are selected. Therefore, we also test the geometric distributions for the 18 events with $E\gtrsim 50$~TeV. We show the $p$-values for all 28 events and the 18 events with $E\gtrsim 50$~TeV in Table~\ref{tab:pvalues}. One can see that the $p$-values are fairly insensitive to the energy cut.
\begin{table}[hb!]
 \centering
  \renewcommand{\arraystretch}{1.0}
\begin{tabular}{c|c|c|c}
  \hline
  & $\bar \alpha=0.17$ & $\bar \alpha=0.25$& Homogeneous  \\
  \hline  
  \hline
  all 28 events & 22.0\% & 20.3\%  & 72.1\% \\
  \hline
  18 events with $E\gtrsim 50$~TeV  & 35.5\% & 31.8\%  & 84.2\% \\
  \hline
 21 cascade events & 41.9\% & 38.8\%   & 95.4\% \\
  \hline
\end{tabular}
\caption{The $p$-value's for three different hypothesis's using all the events, only the events with $E\gtrsim 50$~TeV and only the cascade events.}
  \label{tab:pvalues}
\end{table}
In the last row of Table~\ref{tab:pvalues}, we also show the $p$-values for only the cascade events considering the fact that the track events could have an origin from the atmospheric muon background. From Table~\ref{tab:pvalues}, one can already see that there is no dramatical difference between $\bar{\alpha}=0.25$ and $\bar{\alpha}=0.17$ cases. This is due to the poor angular resolution of cascade events such that the peaked center of the DM profiles can not be resolved. The increase of the $p$-values for the homogeneous distribution from all 28 events to 21 cascade events is due to the extremely good resolution of the 7 track events.

\begin{figure}[th!]
  \begin{center}
    \hspace*{-0.0cm}
    \includegraphics[width=0.48\textwidth]{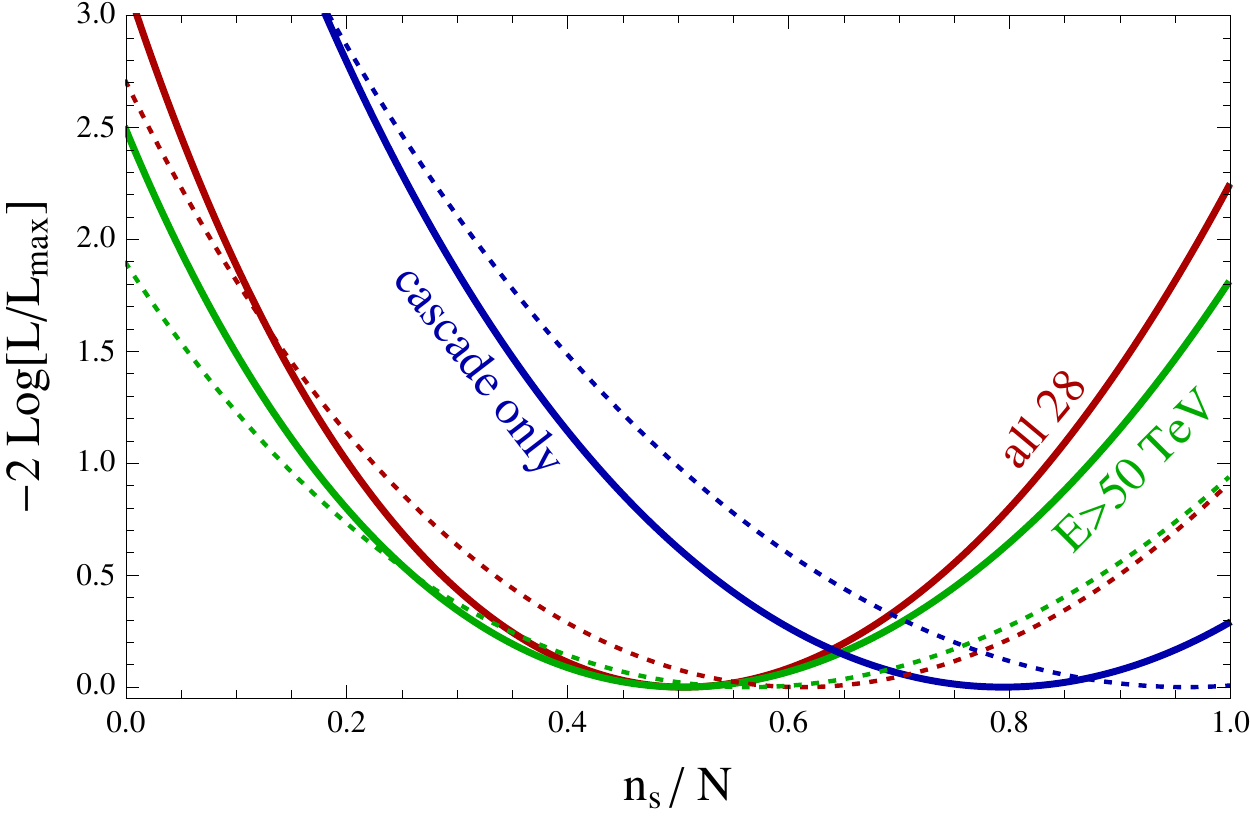} 
    \caption{The log-likelihood as a function of the signal event fraction for all 28 events, 18 events with $E\gtrsim 50$~TeV and 21 cascade events. The solid lines are for $\bar{\alpha}=0.17$ and the dotted lines are for $\bar{\alpha}=0.40$. 
}    \label{fig:log-likelihood-distribution}
  \end{center}
\end{figure}

\paragraph{\bf Geometric analysis using the likelihood ratio test}
\label{sec:geometry_like}
In this section we check the compatibility of the data with the DM
profile using a likelihood ratio test, which was used by the IceCube collaboration in their point source analysis~\cite{Braun:2008bg}. We first treat the homogenous distribution as the null hypothesis with an alternative homogeneous plus DM hypothesis. We define the likelihood function as
\beqa
\mathcal{L}(n_s)=\prod_{i}^{N} \left[ n_s\,\mathcal{S}_i + (1-\frac{n_s}{N})\,\mathcal{B}_i   \right] \,,
\label{eq:log-likelihood-definition}
\eeqa
where $\mathcal{B}_i$ is the homogeneous background contribution and $\mathcal{S}_i$
is the signal DM contribution.~\footnote{We have also calculated the $p$-value's for the point source signal hypothesis and found a good agreement with the result obtained by the IceCube collaboration~\cite{IceCube:science}.} The $n_s$ is the number of signal events and will be used to maximize the likelihood. We use the observed data locations and errors convoluted by the DM probability distribution to calculate the signal contribution $\mathcal{S}_i$ as
\beqa
      \mathcal{S}_i=\int \frac{1}{2\pi
        \sigma_i^2}e^{-\frac{|\vec{x}_i - \vec{x}_s|}{2\sigma_i^2}}p_{\rm DM}(\vec{x}_s) d^2\vec{x}_s \,.
\eeqa
Here, $\vec{x}_i$ is a vector, defined in the $(\delta, \alpha)$ plane, from the location of the observed event and $\sigma_i$  is the corresponding angular error. We show the log-likelihood function as a function of signal strength $n_s/N$ in Fig.~\ref{fig:log-likelihood-distribution} for three cases: all 28 events, 18 events with $E\gtrsim 50$~TeV and 21 cascade events. We can see from Fig.~\ref{fig:log-likelihood-distribution}  that the preferred values of $n_s$ are positive, which suggests that a combination of DM plus homogenous distributions fit the data better than the homogenous-only fit. Comparing the best fitted values of $n_s$ for $\bar{\alpha}=0.17$ and $\bar{\alpha}=0.40$, one can see that a larger value of $\bar{\alpha}$ or a flatter DM profile prefers more DM signal events. 

\begin{figure}[th!]
\begin{center}
    \includegraphics[width=0.48\textwidth]{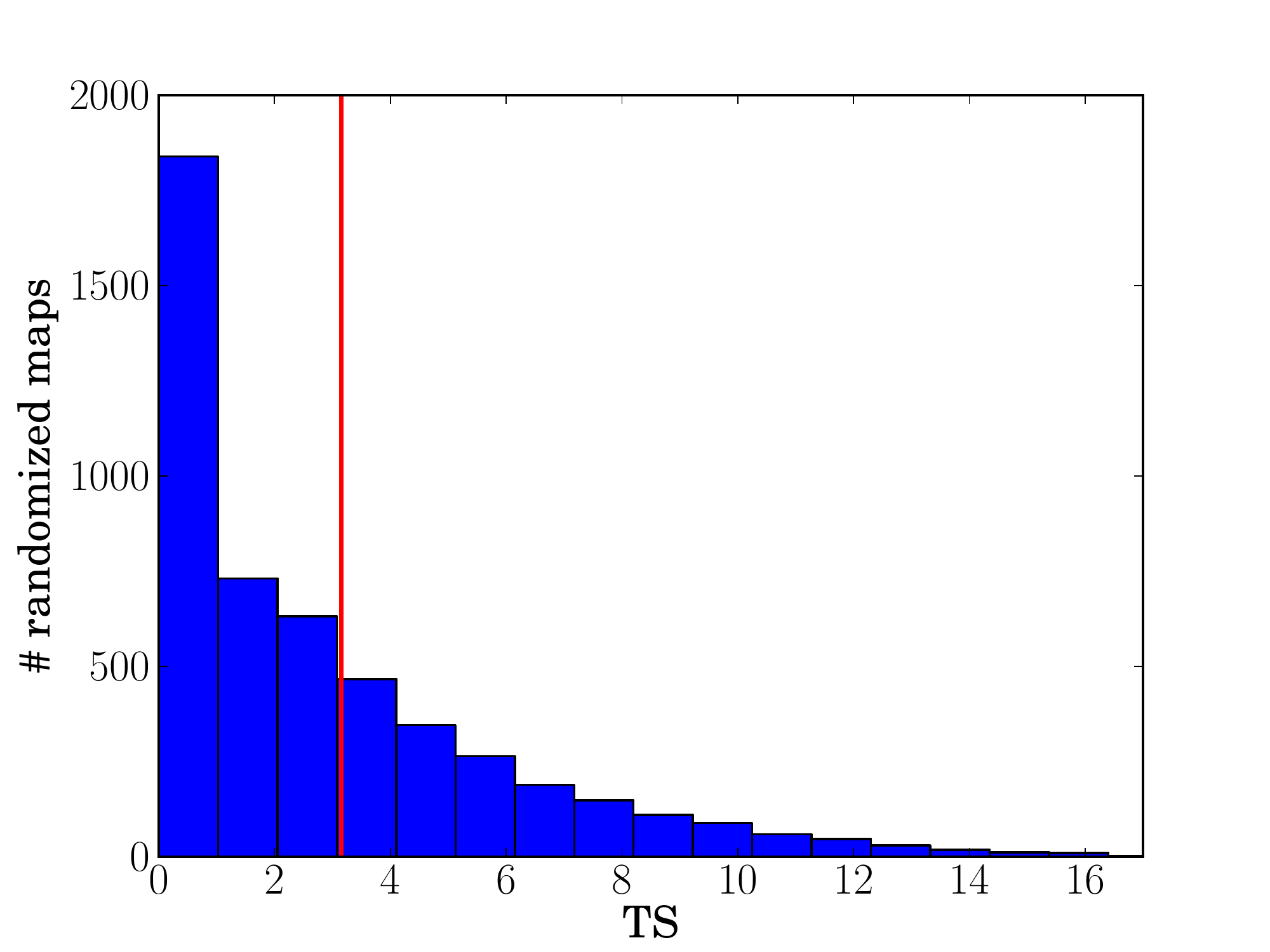}
\caption{The TS distribution for the Einasto model with $\bar
  \alpha=0.17$ ($p_{\rm value}=33.4\%$), the red line is the TS for all 28 events. Here, we have $\mathcal{S}$=DM and $\mathcal{B}$=homogeneous, to have the homogeneous distribution as the null hypothesis. 
}
\label{fig:fit_like}
\end{center}
\end{figure}

To quantify the $p$-value of the data to reject the homogenous-only hypothesis against the homogenous plus DM hypothesis, we calculate the test statistic as 
\beqa
  \mbox{TS}=\max_{n_s} \left\{ 2\log\left[\frac{\mathcal{L}(n_s)}{\mathcal{L}(0)} \right]\right\}\,.
\eeqa
As we did in the last section we compute the $p$-values by scrambling the
events in right ascension angle $\alpha$ with a distribution consistent with the background. For all the 28 IceCube events, we show the histogram for the TS distribution in Fig.~\ref{fig:fit_like} and have the vertical and red line at the real data location. 
For all three choices of events, we show the $p$-values and the $n_s$'s at the maximum 
likelihood for two different values of $\bar{\alpha}$ in Table~\ref{tab:pvalues_like}, which clearly show that the 21 cascade events have a smaller value for the homogenous-only hypothesis. 
\begin{table}[hb!]
 \centering
  \renewcommand{\arraystretch}{1.2}
\begin{tabular}{c|c|c}
  \hline
  & $\bar \alpha=0.17$ & $\bar \alpha=0.25$ \\
  \hline  
  \hline
  all 28 events  & 33.4\% (14.2)  & 36.0\% (15.3) \\
  \hline
  18 events with $E\gtrsim 50$~TeV & 25.0\% (9.1) & 27.2\% (9.5) \\
  \hline
 21 cascade events & 15.8\% (16.7) & 17.9\% (18.0) \\
  \hline
\end{tabular}
  \caption{The $p$-values using the likelihood method using all the events, 
    only the events with $E\gtrsim 50$~TeV and only the cascade events. The numbers in the parenthesis are the numbers of signal events after maximize the log-likelihood. Here, we have $\mathcal{S}$=DM and $\mathcal{B}$=homogeneous in Eq.~(\ref{eq:log-likelihood-definition}).}
  \label{tab:pvalues_like}
\end{table}

\begin{figure}[th!]
  \begin{center}
    \hspace*{-0.0cm}
    \includegraphics[width=0.48\textwidth]{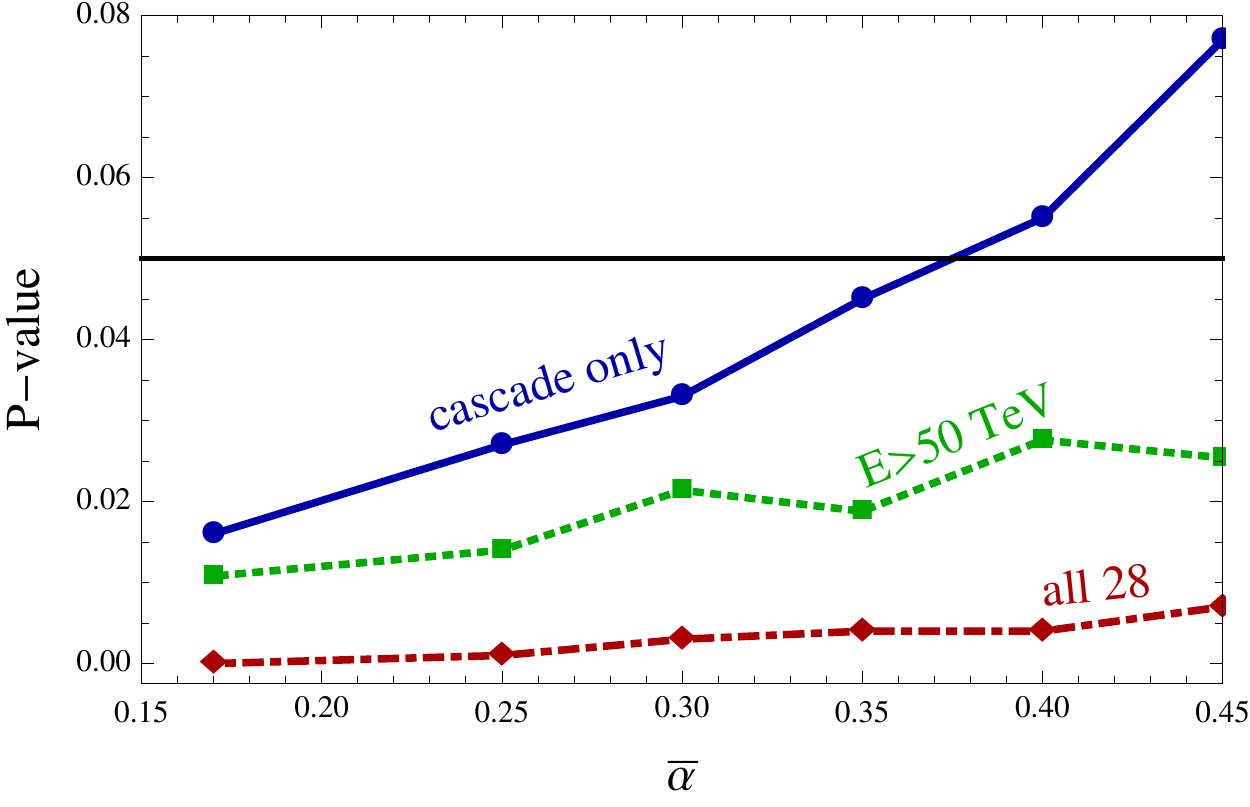} 
    \caption{The $p$-values as a function of $\bar{\alpha}$ of the Einasto DM profiles. A suggestive $p$-value of 0.05 to exclude a certain DM model is shown in the horizontal and black line.  Here, we have $\mathcal{S}$=homogeneous and $\mathcal{B}$=DM, to have the DM distribution as the null hypothesis. 
}    \label{fig:pvalue-alpha}
  \end{center}
\end{figure}

Another interesting question one can ask is whether one can exclude the purely galactic DM hypothesis against the galactic DM plus homogeneous distribution (a part of the homogeneous distribution could come from extragalactic DM or other astrophysical objects). We also calculate the $p$-values for this case. Specifically, one just need to choose $\mathcal{B}$=DM and $\mathcal{S}$=homogeneous in Eq.~(\ref{eq:log-likelihood-definition}) as well as to scramble in terms of the DM profile when we calculate the TS distribution. We show the results in Fig.~\ref{fig:pvalue-alpha} for different values of $\bar{\alpha}$, which clearly show that the pure galactic DM explanation for the data is not preferred for a wide range of $\bar{\alpha}$. For the 21 cascade events and for a flatter DM profile with a larger $\bar{\alpha}$, there is still a non-negligible Type-I error for rejecting the pure galactic DM explanation. We have also checked and found that the IceCube data can not exclude the pure galactic DM explanation with an isothermal DM profile, $\rho_{\rm DM}(r) = \rho_0/(1+r^2/r_c^2)$, with a core radius of $r_c=1$~kpc~\cite{Isothermal}. 

%

\paragraph{\bf Neutrino spectra from dark matter decays}
\label{sec:dark-matter-decay}
The energy spectrum of the IceCube neutrino excess has interesting features~\cite{IceCube:science}. First, there are two isolated events at around 1 PeV~\cite{Aartsen:2013bka} with one at $1.04\pm0.16$~PeV and the other one at $1.14\pm0.17$~PeV. Secondly, there is an potential energy cutoff at $1.6^{+1.5}_{-0.4}$~PeV. Thirdly, there is an energy gap or no neutrino events observed in the energy range of $\sim (0.3, 1)$~PeV, which is not significant at this moment.
 Although a wide range of the energy spectrum can be fit by an $E^{-2}$ feature~\cite{IceCube:science}, it is still interesting to explore potential DM produced spectra from particle physics.

To fit the observed spectrum at IceCube, one also needs to consider different detector acceptances at different energies. For different flavors of neutrinos, the acceptance areas vary a lot with the largest one for the electron neutrino. In our analysis below, we don't distinguish different flavors of neutrinos and use the averaged acceptance areas in terms of flavors and declination angles~\cite{IceCube:science}, which are only slightly different from Ref.~\cite{Anchordoqui:2013qsi}. 
Because the uncertainties on the acceptance areas and the large statistical errors, the current IceCube data is not sufficient to distinguish spectra among different particle physics models. So, we consider several representative decaying DM models and study their fit to the observed energy spectrum. We consider candidate models according to the operator dimensions of DM coupling to SM particles.  

\begin{figure}[th!]
\begin{center}
\includegraphics[width=0.48\textwidth]{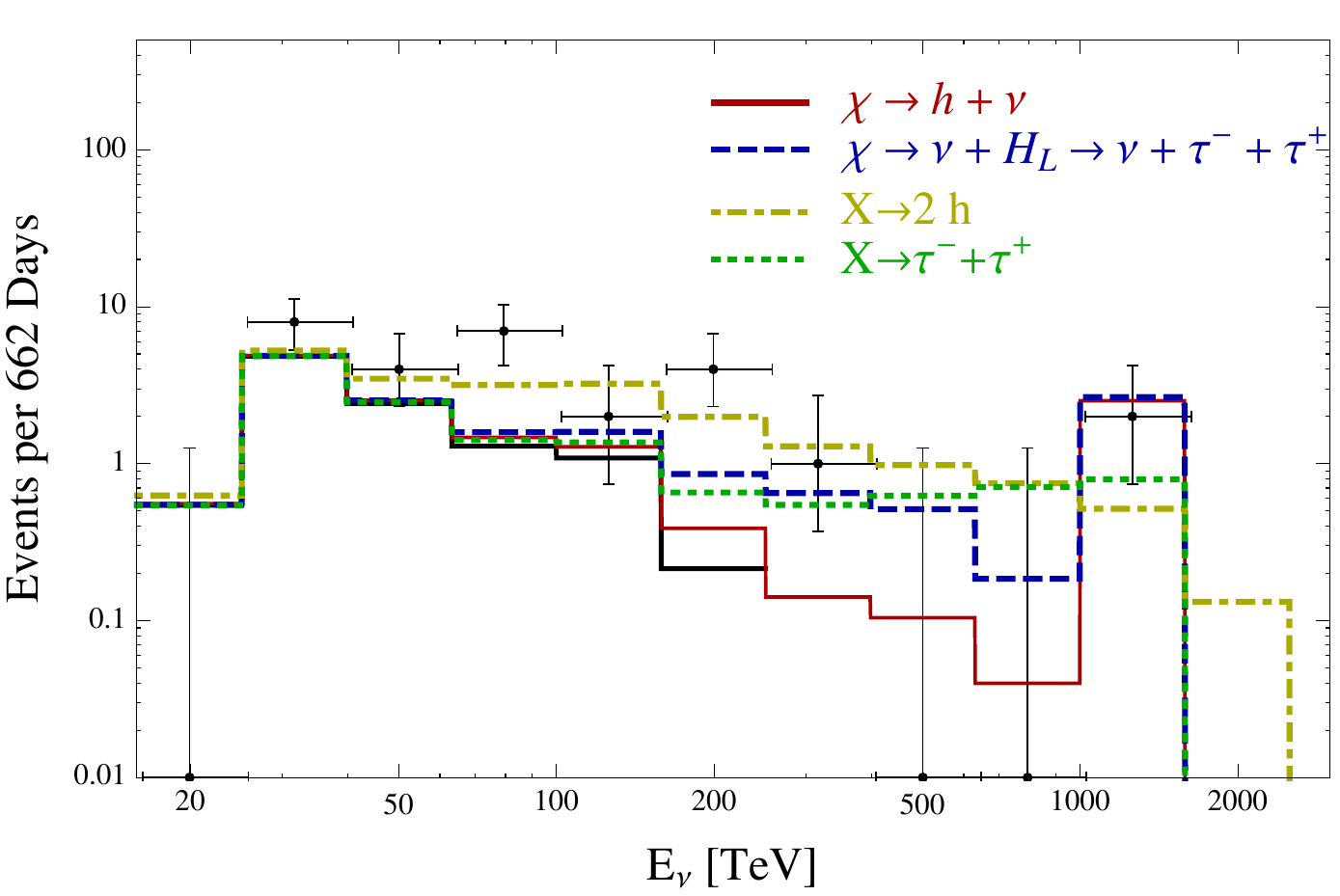}
\caption{The fitted spectra for several DM decay channels. The black and solid line is the atmospheric backgrounds~\cite{Honda:2006qj, Enberg:2008te}. For the two fermion DM cases, the DM mass is  2.2 PeV and both lifetimes are $\tau_{\chi} = 3.5\times 10^{29}$~s. For the two scalar DM cases, the DM mass is 5 PeV and the lifetimes are  $9.2\times 10^{28}$~s and $4.6\times 10^{29}$~s, for $2h$ and $\tau^- + \tau^+$ channels, respectively.
}
\label{fig:spectrum-fit}
\end{center}
\end{figure}

At the renormalizable level and for a fermion DM $\chi$, we consider the operator $\lambda \tilde{H} \bar{L}_L\,\chi$ for DM coupling to the Higgs field in the SM or $\lambda H_L \bar{L}_L\chi$ in the lepton-specific two-Higgs doublet models, which has DM decays as $\chi \rightarrow h + \nu$ and $\chi \rightarrow \nu + H_L \rightarrow \nu + \tau^+ + \tau^-$, respectively. Fixing the fermion DM mass to 2.2 PeV, we show the fitted spectra in Fig.~\ref{fig:spectrum-fit} after using \texttt{PYTHIA}~\cite{Sjostrand:2007gs} for SM particles decay and hadronization. We sum the experimental error and systematical background error in quadrature to calculate the total chi-square for the goodness of fit.  For the two fermion DM decay spectra, a dip feature exists because of the combination of mono-energetic and continuous neutrinos. For a scalar DM, one can have the renormalizable coupling to the SM Higgs boson as simple as $\mu\,XHH^\dagger$, which simply mediates the decay of $X \rightarrow 2h$. Beyond the renormalizable level, one could have DM mainly couple to two leptons via $\epsilon\,m_\tau X \tau^+ \tau^-/\Lambda$, so the decay channel is $X \rightarrow \tau^+ \tau^-$. Fixing the scalar DM mass to be 5 PeV, we also show the fitted spectra in Fig.~\ref{fig:spectrum-fit} (see~\cite{Feldstein:2013kka,Esmaili:2013gha} for other spectra from DM decays).


\paragraph{\bf Conclusions and discussion}
\label{sec:conclusion}
Our geometrical analysis has already shown that a combination of the galactic DM contribution and a homogenous spectrum, which could be due to additional extragalactic sources, provides the best fit to the data. A purely galactic DM origin for the 28 events is not preferred unless a flatter DM spacial profile like the isothermal one is used. The IceCube has more data to be analyzed and collected, so a more robust conclusion can be drawn in the coming years. Other than IceCube, another neutrino telescope, ANTARES~\cite{AdrianMartinez:2012rp}, has reached a comparable sensitivity in some declination angle region. A geometric test for the compatibility between the neutrinos  (excess) observed in ANTARES and a decaying DM will be demanding.

\begin{figure}[th!]
    \begin{center}
        \includegraphics[width=0.48\textwidth]{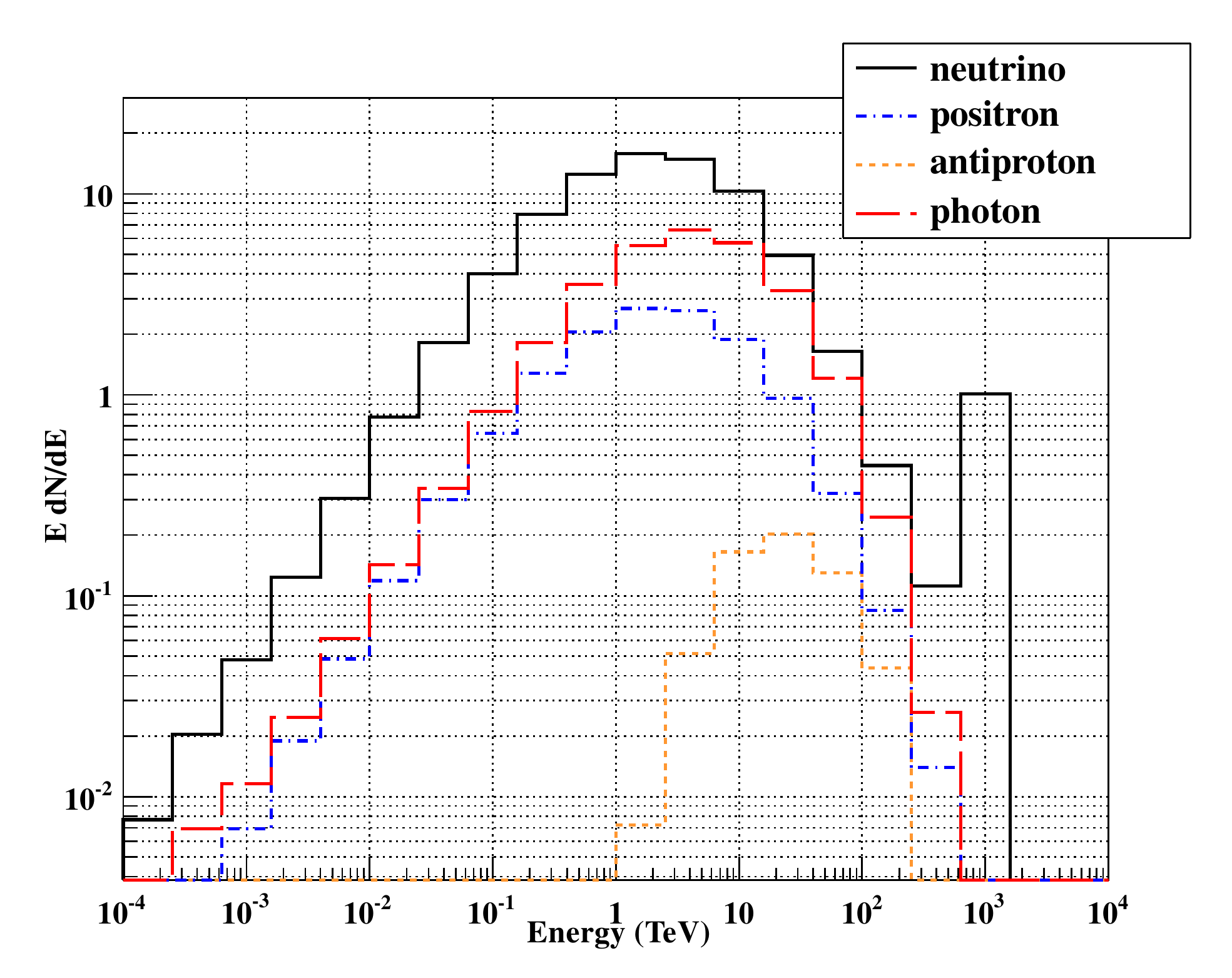}
    \end{center}
    \caption{Neutrino, positron, antiproton and photon yields from a single DM decay with a 2.2 PeV mass and the decay channel $\chi \rightarrow h + \nu$.}
    \label{fig:yield}
\end{figure}

Beyond the neutrino signal from DM, one could also search for other correlated and for sure model-dependent cosmic ray signatures from the DM decays at other experiments like Fermi LAT~\cite{Ackermann:2012rg,Abdo:2010ex}, PAMELA~\cite{Adriani:2008zr,Adriani:2010rc}, AMS-02~\cite{Aguilar:2013qda} and HESS~\cite{Aguilar:2013qda}. In the few respective models considered in Fig.~\ref{fig:spectrum-fit}, additional photons, positrons and antiprotons can be produced at the same time when a neutrino signal is generated. Using the model with $\chi \rightarrow h + \nu$ as an example, we show the yields of neutrino, positron, antiproton and photon from a single DM decay in Fig.~\ref{fig:yield}. One can see that the neutrino yield is considerably higher than the photon, positron and antiproton yields in every bin. Furthermore, because of the long DM lifetime of $10^{28}-10^{29}$~s, the predicted photon, positron and antiproton fluxes have been checked to satisfy the current cosmic ray constraints.

The PeV scale DM considered here is definitely beyond the scope of high energy collider searches. If additional interactions exist between DM and quarks, the direct detection experiments may see a signature~\cite{Albuquerque:2003ei}. If the IceCube excess is indeed due to decaying DM, a new avenue to understanding the DM properties will be opened.

\paragraph{\bf Acknowledgments}
We would like to thank Vernon Barger, Daniel Chung, Francis Halzen,  Claudio Kopper, Naoko Neilson and Nathan Whitehorn for useful discussion. The work is supported by the U. S. Department of Energy under the contract DE-FG-02-95ER40896. YB is partially supported by startup funds from the UW-Madison. The work of RL is partially supported by U. S. Department of Energy under the contract DE-FG-02-95ER40899 and by the Michigan Center for Theoretical Physics. JS acknowledges support from the Wisconsin IceCube Particle Astrophysics Center.

\bibliography{Icecube28}
\bibliographystyle{apsrev}
 \end{document}